\documentclass[%
 preprint,
 amsmath,amssymb,
 aps,floatfix,superscriptaddress]{revtex4-2}

\usepackage{lipsum}
\usepackage{natbib}
\usepackage{url}
\usepackage[dvipsnames]{xcolor}
\usepackage{graphicx}
\usepackage{dcolumn}
\usepackage{bm}
\usepackage{siunitx}

\makeatletter
\newcommand{\newparallel}{\mathrel{\mathpalette\new@parallel\relax}}
\newcommand{\new@parallel}[2]{%
  \begingroup
  \sbox\z@{$#1T$}
  \resizebox{!}{\ht\z@}{\raisebox{\depth}{$\m@th#1/\mkern-5mu/$}}%
  \endgroup
}

\usepackage{mathtools}

\DeclarePairedDelimiter\ket{\lvert}{\rangle}

\newcommand{\beginsupplement}{%
        \setcounter{table}{0}
        \renewcommand{\thetable}{S\arabic{table}}%
        \setcounter{figure}{0}
        \renewcommand{\thefigure}{S\arabic{figure}}%
        \renewcommand{\thesubsection}{\roman{subsection}}
     }
     
\usepackage[titles]{tocloft}
\begin{document}

\preprint{}

\title{Anisotropic Laser-Pulse-Induced Magnetization Dynamics in van der Waals Magnet Fe$_3$GeTe$_2$}

\author{Tom Lichtenberg}
  \email{t.lichtenberg@tue.nl}
\author{Casper F. Schippers}%
\author{Sjoerd C.P. van Kooten}%
\author{Stijn G.F. Evers}%
\affiliation{
 Department of Applied Physics, Eindhoven University of Technology, P.O. Box 513, 5600 MB Eindhoven, The Netherlands}%
 \author{Beatriz Barcones}
 \affiliation{NanoLabTUe, Eindhoven University of Technology, P.O. Box 513, 5600 MB Eindhoven, The Netherlands}
\author{Marcos H. D. Guimar\~{a}es}
\affiliation{Zernike Institute for Advanced Materials, University of Groningen, \\ Nijenborgh 4, 9747 AG Groningen, The Netherlands
}%
\author{Bert Koopmans}%
\affiliation{
 Department of Applied Physics, Eindhoven University of Technology, P.O. Box 513, 5600 MB Eindhoven, The Netherlands}%

\date{\today}

\begin{abstract}

\noindent Femtosecond laser-pulse excitation provides an energy efficient and fast way to control magnetization at the nanoscale, providing great potential for ultrafast next-generation data manipulation and nonvolatile storage devices.
Ferromagnetic van der Waals materials have garnered much attention over the past few years due to their low dimensionality, excellent magnetic properties, and large response to external stimuli.
Nonetheless, their behaviour upon fs laser-pulse excitation remains largely unexplored.
Here, we investigate the ultrafast magnetization dynamics of a thin flake of Fe$_3$GeTe$_2$ (FGT) and extract its intrinsic magnetic properties using a microscopic framework.
We find that our data is well described by our modelling, with FGT undergoing a slow two-step demagnetization, and we experimentally extract the spin-relaxation timescale as a function of temperature, magnetic field and excitation fluence.
Our observations indicate a large spin-flip probability in agreement with a theoretically expected large spin-orbit coupling, as well as a weak interlayer exchange coupling.
The spin-flip probability is found to increase when the magnetization is pulled away from its quantization axis, opening doors to an external control over the spins in this material.
Our results provide a deeper understanding of the dynamics van der Waals materials upon fs laser-pulse excitation, paving the way towards two-dimensional materials-based ultrafast spintronics.
\end{abstract}

\maketitle
\tableofcontents
\newpage
\section{Introduction}
\noindent As conventional data storage and manipulation technologies are reaching their fundamental limits in terms of bit density and processing speed, the need for faster and more efficient solutions has never been more apparent.
The possibility of sub-ps control of magnetization with laser-pulses was first discovered in 1996 by Beaurepaire \emph{et al.}, and has been a major subject of research since then~\cite{beaurepaire1996ultrafast}.
Large strides towards applications that operate on this ultimate timescale have been made with the discovery of fs-laser induced all-optical magnetization switching (AOS)~\cite{stanciu2007all}, ultrafast spin-current generation~\cite{Malinowski2008ControlMomentum, Choi2014SpinDemagnetization, Schellekens2014UltrafastExcitation}, and spin-wave excitation~\cite{van2002all, satoh2006coherent, tzschaschel2019tracking}.
Simultaneously, several theoretical frameworks have been developed to understand the microscopic mechanisms governing this phenomenon, both in terms of local~\cite{zhang2000laser, koopmans2005unifying, djordjevic2007connecting} and non-local~\cite{Battiato2010SuperdiffusiveDemagnetization} angular momentum dissipation.

The recent discovery of long-range magnetic order in atomically-thin van der Waals (vdW) materials~\cite{lee2016ising, deng2018gate, gong2017discovery, huang2017layer, bonilla2018strong} provides a new and exciting platform for ultrafast spintronics.
Their ability to stack without the need for lattice matching makes them ideal candidates for multi-component applications based on for instance spin-current injection, spin-wave excitation and AOS.
The first step towards such applications is studying the underlying fundamental phenomena in vdW magnets, namely laser-induced demagnetization and spin precession.
Additionally, their ultrafast magnetization dynamics can unveil some of their important magnetic and spintronic properties, such as spin-flip rates and electron-phonon coupling.
Although magnetic vdW materials have been studied extensively over the past few years, their magnetic response to ultra-short fs laser-pulses remains scarcely explored~\cite{zhang2020laser, sun2021ultra}.

One of the most promising vdW ferromagnets is Fe$_3$GeTe$_2$ (FGT - Figure\,\ref{fig:1}a), due to its relatively high Curie temperature ($T_\text{C}$) of $220-230$ K and strong perpendicular magnetic anisotropy~\cite{fei2018two}.
This makes FGT especially suitable for application in spin-valves~\cite{wang2018tunneling, li2021large} and allows for efficient spin-orbit torques~\cite{alghamdi2019highly, shin2021spin, zhao2021van}. Furthermore, its properties allow for the stabilization of complex spin structures such as spin spirals and skyrmions~\cite{wu2020neel, meijer2020chiral}.
Unraveling the physics governing laser-pulse induced demagnetization dynamics in FGT brings us one step further towards unifying the fields of vdW materials, spintronics and (ultrafast) photonics.

In this work, we study the fs laser-pulse induced demagnetization dynamics of thin-film FGT.
Specifically, we measure the timescales involved with the demagnetization process as a function of laser fluence, ambient temperature and applied magnetic field.
We compare our experimental results to calculations based on the well-established microscopic 3-temperature model (M3TM)~\cite{koopmans2010explaining}, and find very good quantitative agreement.
We find that FGT undergoes a two-step demagnetization (type-II), in agreement with an earlier study~\cite{sun2021ultra}, which we attributed to its low Curie temperature and relatively large atomic moment.
Nonetheless, our theoretical analysis indicates a surprisingly large spin-flip probability, contrary to the earlier claims of long spin lifetimes, but in agreement with theoretical expectations of large spin-orbit coupling in FGT.
The laser-induced demagnetization is found to be strongly dependent on an applied in-plane magnetic field, consistent with an anisotropy in the spin-flip probability, which has been predicted for other material classes~\cite{zimmermann2012anisotropy, zimmermann2016fermi}.
Moreover, contrary to recent studies in Cr$_2$Ge$_2$Te$_6$ (CGT)~\cite{sun2021ultra}, we were unable to excite ferromagnetic resonance (FMR) modes.
We hypothesize that this is caused by strong dephasing of the FMR mode due to weak interlayer coupling and in-plane anisotropy variations.

\section{Methods and sample characterization}

\begin{figure}[t!]
\centering
\includegraphics[width=10cm]{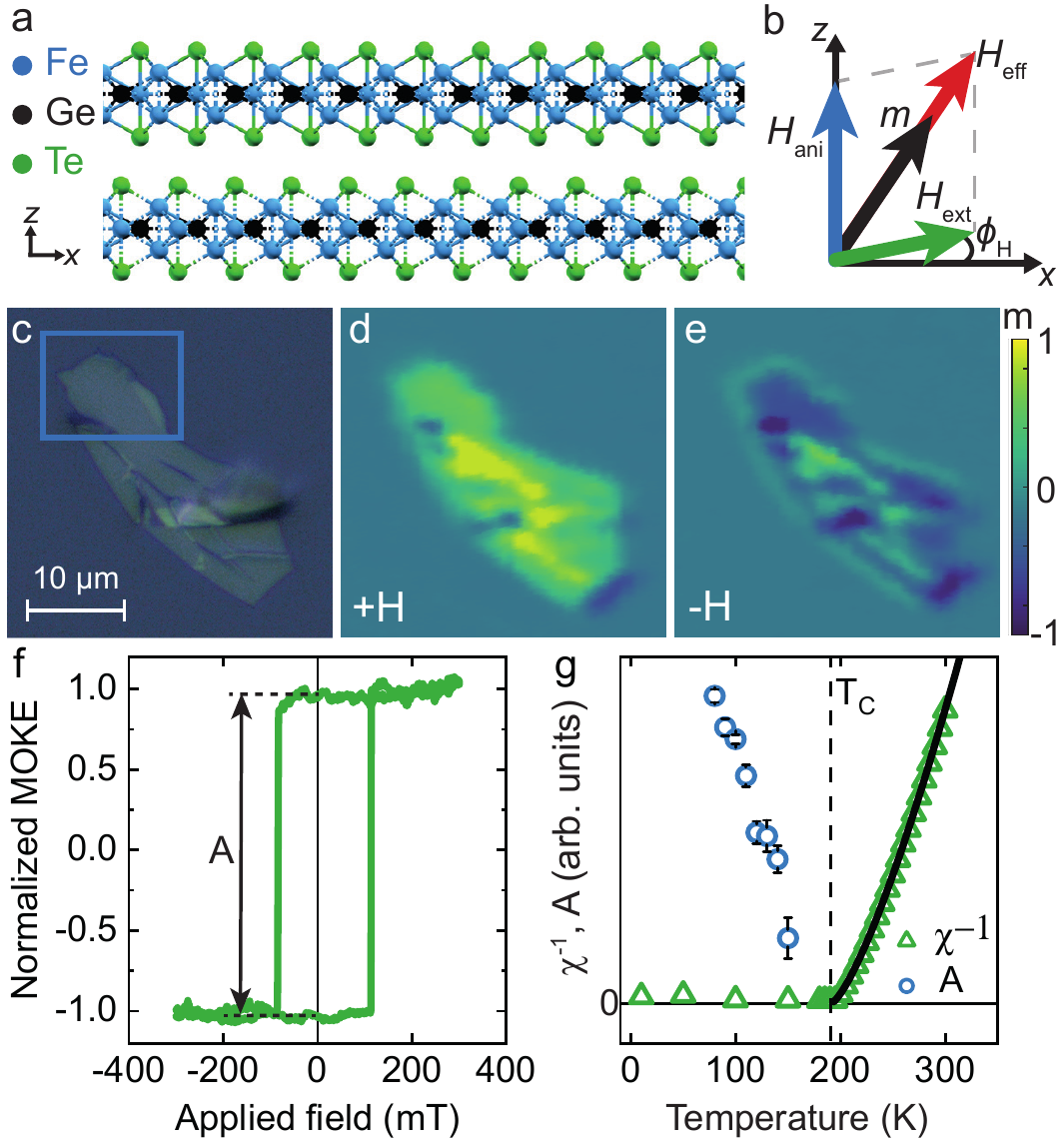}
\caption{\label{fig:1} (a) Crystal structure of FGT, drawn using \emph{Mercury}~\cite{macrae2020mercury} (b) Schematic representation of the orientation of the magnetization and relevant (effective) magnetic fields. $\phi_\text{H}$ indicates the angle between the applied field and the sample plane ($x$-axis) (c) Optical micrograph of the studied FGT flake. The blue square indicates the homogeneous area on which experiments are conducted. (d,e) MOKE imaging of the flake after saturation with a field of +400 mT and -400 mT, respectively. (f) Typical MOKE hysteresis loop measured on the FGT flake for $\phi_\text{H}\!=\!19^{\circ}$. (g) Temperature dependence of the inverse magnetic susceptibility of the bulk crystal (green triangles) and the MOKE signal measured on the flake (blue circles), as defined in figure\,f. The black line indicates the fit with a power law, used to extract the Curie temperature.  Measurements d, e and f are done at 80 K.}
\end{figure}

\noindent Our samples are fabricated through mechanical exfoliation of commercially available FGT crystals (HQ Graphene) in inert nitrogen atmosphere and transferred to an optical cryostat without air exposure, where the sample is then kept in high vacuum ($<10^{-6}$ mBar) throughout all experiments.
To measure the magneto-optical response we apply an external magnetic field ($H_\text{ext}$) at an angle $\phi_\text{H}$ with respect to the sample plane.
Due to the out-of-plane magnetic anisotropy field of FGT ($H_\text{ani}$), the magnetization $m$ will be tilted from the out-of-plane direction, according to the effective magnetic field $H_\text{eff}$.
Figure\,\ref{fig:1}b shows a diagram indicating all relevant directions and angles.

After selecting a thin FGT flake (figure\,\ref{fig:1}c) using an optical microscope, the flake is probed using magneto-optical Kerr effect (MOKE) microscopy at low temperatures ($T = 80$K). This way, magnetically homogeneous areas with clear single-domain behavior of several \SI{}{\micro\metre}$^2$ are identified, figure\,\ref{fig:1}d and e.
The flake for which the results are shown here was characterized by atomic force microscopy after all other experiments were completed to obtain its thickness, $t=16.8\pm0.7$ nm, corresponding to $20\pm1$ monolayers (See section\,V of the supplementary information~\cite{supplement}).

For our MOKE measurements we use a mode-locked Ti:Sapphire laser centered at 735 nm with a repetition rate of 82 MHz and 85 fs of pulse width at the sample location, with a spot size of $2.9\times 1.7$ \SI{}{\micro\metre}$^2$.
Using hysteresis curves of MOKE signal versus $H_\text{ext}$ at $\phi_\text{H}$ = 19 $^\circ$ (figure\,\ref{fig:1}f), we obtain the MOKE amplitude $A$ versus temperature (blue circles in figure\,\ref{fig:1}g).
We find that the MOKE signal decreases abruptly around 200 K, in agreement with vibrating sample magnetometry measurements done in a bulk FGT flake (figure\,\ref{fig:1}g, green triangles).
The inverse magnetic susceptibility ($\chi^{-1}$) of the bulk crystal is well described by a power law for the paramagnetic phase given by $(T-T_{\text{C}})^\gamma$.
A clear transition point corresponding to the Curie temperature $T_\text{C} = 191 \pm 2$ K is observed, with $\gamma=1.3 \pm 0.1$, which is close to the typical exponent for a three-dimensional Ising magnet ($\gamma\sim1.25$)~\cite{guida19973d}.
We note that the Curie temperature is significantly lower than expected for stoichiometric FGT~\cite{fei2018two, deiseroth2006fe3gete2, chen2013magnetic, zhu2016electronic, zhuang2016strong}, indicating that our material is slightly Fe deficient~\cite{may2016magnetic}.
This is further confirmed by Energy-dispersive X-ray spectroscopy (EDX) measurements (see section\,IV of the supplementary information~\cite{supplement}).

The laser-induced demagnetization measurements described below are performed by pump-probe spectroscopy using the same setup as above.
The higher intensity pump beam, with focused area of $9.6\times 6.6$ \SI{}{\micro\metre}$^2$, is obtained from the same laser and overlapped onto the probe spot.
For additional experimental details see section\, I of the supplementary information~\cite{supplement}.

\section{Experimental results and Discussion}
\subsection{Demagnetization dynamics of FGT}

\begin{figure}[t!]
\centering
\includegraphics[width=10cm]{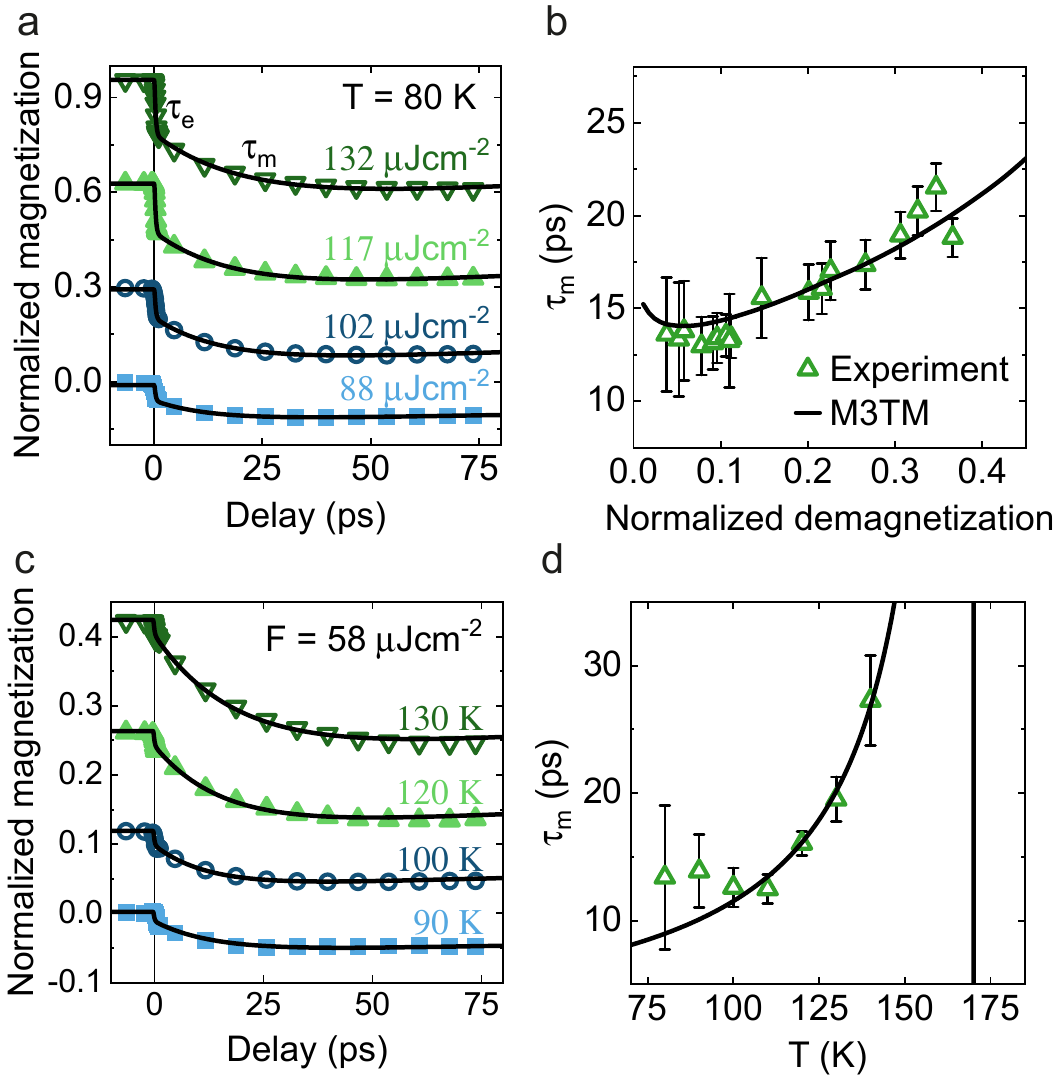}
\caption{\label{fig:2} (a) Laser-induced demagnetization measured at four different laser fluences. The data fitting (black line) was done using the procedure described in Sec\,III of the supplementary information~\cite{supplement}. (b) Magnetic timescale $\tau_\text{m}$ as a function of the normalized total demagnetization. (c) Laser-induced demagnetization for different ambient temperatures. (d) Magnetic timescale $\tau_\text{m}$ as a function of temperature, fitted with equation\,\ref{eq3} (black line). The data in a and c are normalized using the methods described in section\,II of the supplementary information and offset vertically for clarity~\cite{supplement}.}
\end{figure}

\noindent As can be seen in figure\,\ref{fig:2}a, a clear two-step laser-induced demagnetization process is observed in our samples, indicative of type-II demagnetization, similar to rare-earth ferromagnets Gd and Tb~\cite{vaterlaus1991spin, kim2009ultrafast} as well as the vdW magnets CGT~\cite{zhang2020laser, sun2021ultra}.
The total magnetization quenching is proportional to the pump laser fluence, which is attributed to the increased heating towards the Curie temperature.
Two distinct timescales can be distinguished: $\tau_\text{e}$ and $\tau_\text{m}$. 
They are commonly referred to as the electron-phonon relaxation and magnetic timescale respectively.
How these timescales relate to the physics of laser-induced demagnetization will be discussed later.

To extract the relevant demagnetization timescales, the data are fitted with an analytical solution of the phenomenological 3-temperature model (3TM)~\cite{dalla2007influence}.
The fitting procedure is discussed in more details in section\,III of the supplementary information~\cite{supplement}.
Although some temperature dependence of $\tau_\text{e}$ has been reported in literature~\cite{hodak1998ultrafast, wang2010temperature}, we assume a temperature (and thus fluence) independence in our analysis to avoid overparameterization of the model. Furthermore, this allows us to study the changes of $\tau_\text{m}$, which are typically large~\cite{koopmans2010explaining,roth2012temperature}. 
A global electron-phonon relaxation timescale of $\tau_\text{e} = 794 \pm 23$ fs is found, which is a typical value for ferromagnetic transition metals and alloys based on them~\cite{schoenlein1987femtosecond, sun1993femtosecond, groeneveld1995femtosecond,hodak1998ultrafast, koopmans2010explaining, roth2012temperature}.

 The timescale $\tau_\text{m}$ is extracted from the fluence-dependent curves (Figure\,\ref{fig:2}a) and plotted in figure\,\ref{fig:2}b as a function of the normalized total demagnetization for each laser fluence.
An increase from $\sim\!13$ to $\sim\!20$ ps is observed in the studied fluence range.
The error bars are relatively large compared to the apparent spread in the data due to a significant cross-correlation between the magnetic timescale and the typical timescale that governs heat dissipation in the flake.
Related to its fluence dependence, the magnetic timescale is also dependent on the ambient temperature (figure\,\ref{fig:2}d).
We observe a critical slowing down of the magnetization dynamics as $T$ approaches $T_\text{C}$ which is well reported in literature for standard ferromagnets~\cite{kise2000ultrafast, roth2012temperature, kantner2011determination, kimling2014ultrafast}.

 Detailed information on the dynamic magnetization properties of FGT can be obtained by employing the microscopic 3-Temperature Model (M3TM)~\cite{koopmans2005unifying, koopmans2010explaining}, which has been tested extensively on a wide variety of materials~\cite{koopmans2010explaining, roth2012temperature, iihama2015ultrafast, bobowski2017influence, padmanabhan2020coherent}.
Similar to the phenomenological 3-Temperature model for ultrafast demagnetization, the magnetic system is subdivided into three separate systems which can exchange energy as well as angular momentum: the electron (e), the phonon (p) and the spin (s) system~\cite{beaurepaire1996ultrafast}, each characterized by a temperature, $T_\text{e}$, $T_\text{p}$ and $T_\text{s}$. respectively.
A model Hamiltonian is set up to describe the three systems and their interactions. 
Here, we only focus on the key features of the model; for an extensive treatment, we refer to the original papers~\cite{koopmans2005unifying, koopmans2010explaining}.
Upon laser excitation, the electron system is heated, bringing the total system in a non-equilibrium state.
We assume the subsystems are at internal thermal equilibrium at all times.
Furthermore, the heat capacity of the spin system is considered to be negligible.
Under these assumptions, energy exchange between the electron and phonon systems is governed by the 2-temperature model (2TM)~\cite{anisimov1974electron}.
An exemplary solution is presented in figure\,\ref{fig:3}a, using parameters shown in section\,VIII of the supplementary information~\cite{supplement}.
The laser-pulse significantly heats up the electron system beyond the Curie temperature.
Hereafter, the electron and phonon system equilibrate on a timescale given by $\tau_\text{e}$.

\begin{figure}[t!]
\centering
\includegraphics[width=10cm]{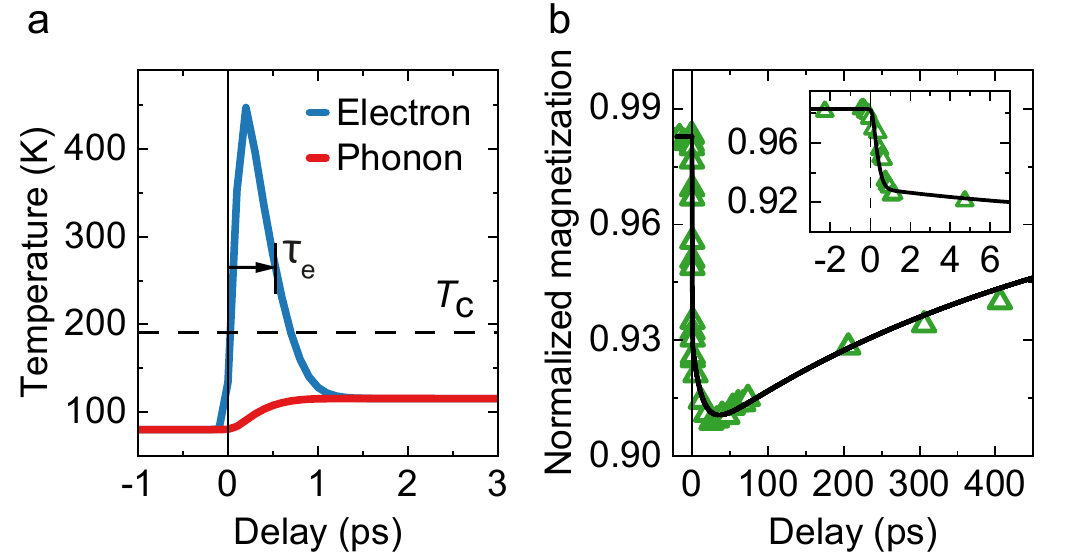}
\caption{\label{fig:3}(a) Calculated electron (blue) and phonon (red) temperatures, using the M3TM. (b) Demagnetization trace (with $F=55$ \SI{}{\micro\joule \meter}$^{-2}$ and T = 80 K), where the black line is the optimal M3TM fit to the data. The inset shows the same data, zoomed in on the first few ps.}
\end{figure}

The angular momentum transfer between the subsystems is represented by $\tau_\text{m}$ and taken to follow the Elliot-Yafett mechanism for spin relaxation~\cite{elliott1954theory, yafet1963solid}.
Here, an electron-phonon scattering event leads to energy exchange between the electron and phonon system.
However, due to spin-orbit coupling, the single-electron state is a mix of spin-up and down down states, of the form $\Psi_{\vec{k}}=a_{\vec{k}} \ket{\uparrow} + b_{\vec{k}} \ket{\downarrow}$~\cite{elliott1954theory, yafet1963solid}.
This leads to spin-mixing of electron states near the Fermi level, quantified by the material dependent spin-mixing parameter $\langle b^2 \rangle$.
Consequently, there is a finite probability $a_\text{sf} \propto \langle b^2 \rangle$ for a spin-flip to occur during an electron-phonon scattering event, thereby transferring angular momentum between the spin system and the lattice.
Transforming the model Hamiltonian to rate equations, the normalized magnetization $m$ can be calculated:~\cite{koopmans2010explaining}: 
\begin{align}
\label{eq2}
\dfrac{\text{d} m}{\text{d} t} =&R \frac{T_\text{p}}{T_\text{C}} \left[1-m\,\text{coth} \left(\frac{T_\text{C}}{T_\text{e}}\right)   \right] \,,\\
\label{eq2b}
R =& \frac{8 a_\text{sf} g_\text{ep} k_\text{B} {T_\text{C}}^2 V_\text{at}}{(\mu_\text{at}/\mu_\text{B}){E_\text{D}}^2}\,.
\end{align} 
Here, $k_\text{B}$, $\mu_\text{B}$ and $\mu_\text{at}$ are the Boltmann constant, Bohr magneton and the atomic moment, respectively.
$V_\text{At}$ and $E_\text{D}$ are the atomic volume and the Debye energy. 
The material parameter $R$ quantifies the demagnetization rate, and is used to determine a material dependent figure of merit $T_\text{C}/\mu_\text{at}$ that provides a simple prediction of the demagnetization rate~\cite{koopmans2010explaining}.

An example demagnetization trace, measured at laser fluence $F=55$ \SI{}{\micro\joule\per\square\meter} and 80 K, is shown in figure\,\ref{fig:3}b.
Model parameters, as presented in section\,VIII of the supplementary information~\cite{supplement}, are extracted from literature, and unknown parameters are chosen such that there is a good agreement between experiment and simulations.
We observe the best correspondence for spin-flip probability $a_\text{sf} = 0.14$.
This is relatively large compared to elementary transition metals~\cite{koopmans2010explaining, roth2012temperature, kuiper2014spin}.
However, Kuiper \emph{et al.} have demonstrated that Co/Pt multilayers have significant larger $a_\text{sf}$ than pure Co due to the proximity of Co to the high SOC Pt~\cite{kuiper2014spin}. 
We argue that this is exactly the case in FGT, where the magnetic moment carrying Fe atoms are closely surrounded by the heavy Te atoms. 
This is corroborated by recent observation of large bulk spin-orbit torques observed in FGT~\cite{johansen2019current, martin2021strong}. 
Although we observe a relatively large $a_\text{sf}$, the demagnetization process is relatively slow.
We attribute this to the sub-room-temperature Curie temperature and large atomic moment, leading to a large figure of merit $T_\text{C}/\mu_\text{at}$ and thus slower demagnetization.

The dependence of $\tau_m$ on laser fluence and temperature is well described by the M3TM.
The simulation results are plotted as a black line in figure\,\ref{fig:2}b and d.
We observe the same upward trend for increasing laser fluence, as well as good quantitative prediction of the demagnetization timescale for realistic model parameters.
This indicates FGT behaves very similarly to elementary transition and Rare-earth magnets.
The theoretical temperature dependence of $\tau_\text{m}$, as derived in section\,VII of the supplementary information~\cite{supplement}, reads: 
\begin{align}
\label{eq3}
\tau_\text{m} = \left(\frac{1}{2R}\right) \left(\frac{1}{1-\frac{T + \Delta T_\text{h}}{T_\text{C}}}\right)\,,
\end{align}
which describes the critical slowing down of the dynamics towards the Curie temperature.
Here, $\Delta T_\text{h}$ is the temperature rise due to the average laser induced heating of the FGT flake at high fluence and temperature.
Note that this relation is derived for temperatures close to $T_\text{C}$, so deviations are expected for low temperatures.
Using the previously determined $T_\text{C}$ of 191 K, the data in figure\,\ref{fig:2}d is fitted using this relation, yielding $R=0.15 \pm 0.01$ ps$^{-1}$.
Using equation\,\ref{eq2b}, a spin-flip probability of $\sim\!0.12$ is calculated, which is similar to values found from our time-dependent simulations, and again confirms the large spin-flip probability found for FGT.
Furthermore, we extract $\Delta T_\text{h}=21 \pm 4$ K from our fitting procedure, which corresponds to a remaining magnetization quenching just before laser excitation of several percents. 
Our approach is further validated in section\,X of the supplementary information, where we estimate the temperature rise using hysteresis loops measured with and without sample pumping at negative time delay~\cite{supplement}. 
The latter yields $\Delta T_\text{h}=16 \pm 7$, in line with the value obtained from fitting equation\,\ref{eq3}.  

\subsection{Magnetic-field dependence}

\begin{figure}[t!]
\centering
\includegraphics[width=10cm]{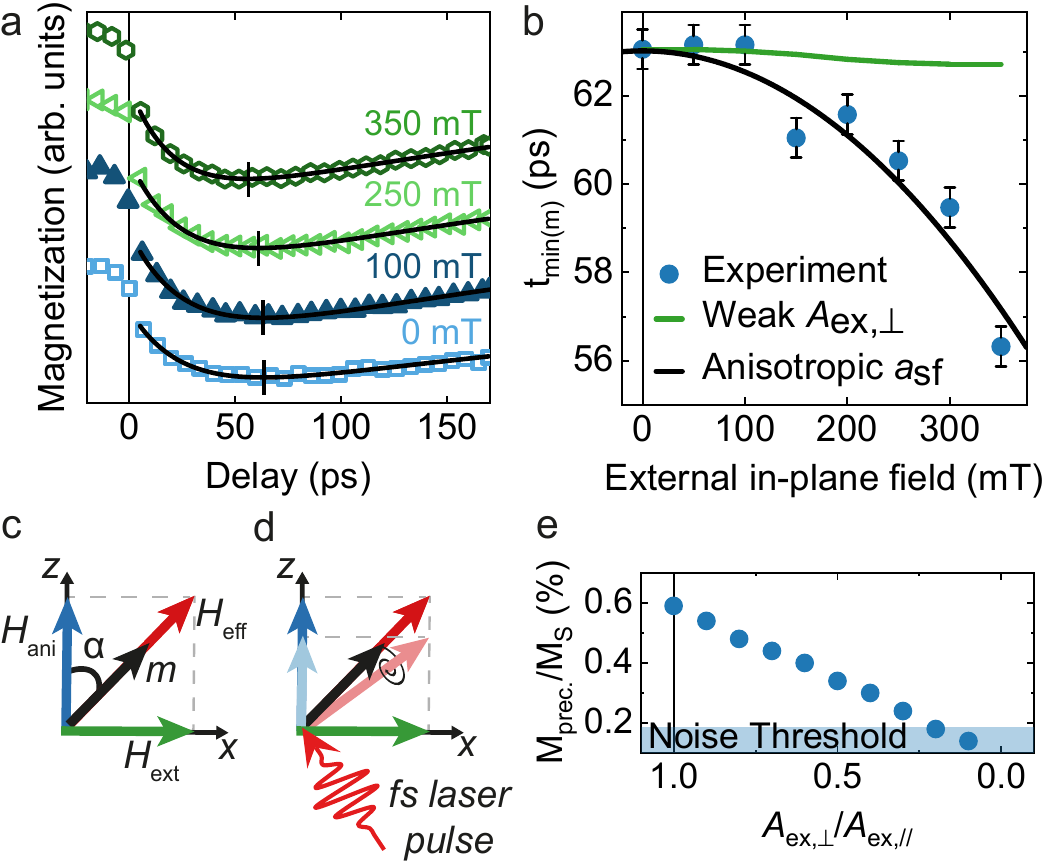}
\caption{\label{fig:4} (a) In-plane ($\phi\!=\!0^\circ$) magnetic field dependence of the demagnetization dynamics. The black line indicate fits with two exponential functions and the horizontal lines indicate the minima of the curves. (b) Delay values of minimum demagnetization as a function of $H_\text{ext}$. The experimental values are shown by blue circles, the green line shows the simulation results using equation\,\ref{LLG} and the black curve is a fit using equation\,\ref{eqani}. (c) Relevant vectors and angles for the field dependence experiments. The light colors indicate changes to the system after excitation. (d) Illustration of the $\Delta K$ mechanism for excitation of FMR modes. (e) Calculated precessional amplitude as a function of the out-of-plane exchange stiffness reduction. The light-blue area indicates our experimental noise threshold. The full datasets are shown in section\,VI of the supplementary information~\cite{supplement}.}
\end{figure}

\noindent Next, in order to study how the symmetry of the FGT crystal structure influences the demagnetization dynamics, we apply an in-plane magnetic field ($\phi_\text{H}=0$) to induce a canting of the magnetization, away from the quantization axis.
The measurements are done at an ambient temperature of 160 K and a laser fluence of $140$ \SI{}{\micro\joule\centi\meter}$^{-2}$ to maximize the demagnetization and thus the effect of the applied field on the magnetization angle (figure\,\ref{fig:4}a). 
We measure a significant speeding up of the laser-induced demagnetization process for increasing in-plane field, which can be associated with an increase of the spin-flip probability.
To visualize the speeding up of the demagnetization dynamics, the delay at the minimum of the demagnetization curve is extracted (figure\,\ref{fig:4}b).
A strong anisotropy in $\langle b^2 \rangle$ and thus the  spin-flip rate is expected in magnetic systems with lowered symmetry and large SOC, of which FGT is a prime example~\cite{zimmermann2012anisotropy}. 
Since the effect of the magnetization canting is symmetric with the applied field, the first order approximation of the spin-flip probability is given by: 
\begin{align}
\label{eqani1}
a_\text{sf} \approx & a_{\text{sf}, 0} (1 + C \alpha^2) \nonumber\\
= & a_{\text{sf}, 0} \left[1 + C \left(\frac{H_\text{ext}}{H_\text{ani}}\right)^2 \right]\,,
\end{align}
where $a_{\text{sf}, 0}$ is the field-free spin-flip probability and $\alpha$ the angle between between the magnetization and the anisotropy field (figure\,\ref{fig:4}c), given by $H_\text{ext}/H_\text{ani}$ for small $\alpha$. 
Here we use $H_\text{ani}=2K_\text{U}/\mu_0 M_\text{S}$ with $K_\text{U}$ the anisotropy constant and $M_\text{S}$ the saturation magnetization (the values used can be found in section\,VIII of the supplementary information~\cite{supplement}).
The phenomenological parameter $C$ is proportional to the ratio between the two orthogonal components of the spin-flip probability, and therefore a measure for its anisotropy.
An increase of $a_{\text{sf}}$ leads to lower $\tau_\text{m}$, which can be expressed in a shift of the minimum of the demagnetization curve $t_\text{min(m)}$.  
The M3TM is used to find the relation between $a_\text{sf}$ and $t_\text{min(m)}$ (See section\,IX of the supplementary information~\cite{supplement}). 
In the lowest order, $t_\text{min(m)}$ scales linearly with $a_\text{sf}$, following:   
\begin{align}
\label{eqani}
t_\text{min(m)} =t_0 - t_1 a_\text{sf} \,,
\end{align}
\noindent where from the M3TM simulations we find $t_0=107.8$ ps and $t_1=324.7$ ps. 
Our results shown in figure\,\ref{fig:4}b are well fitted using the equation above with $C\approx 13$, indicating a (very) strong anisotropy of $a_{\text{sf}}$.
Our results are realistic when compared to calculations for other materials with reduced symmetry~\cite{zimmermann2012anisotropy, zimmermann2016fermi}.
However, in order to gain a deeper microscopic insight on the spin relaxation anisotropy in vdW materials in general and FGT specifically, \emph{ab-initio} calculations have to be carried out for these materials as well.

Another potential motivation for our field-dependent measurements is to excite FMR modes with the $\Delta K$ mechanism, which is schematically shown in figure\,\ref{fig:4}d. Here, a demagnetization reduces the anisotropy of the FGT, thereby quickly reorienting the effective field~\cite{van2002all}.
This leads to the excitation of damped precessional dynamics.
Even though FMR modes have been shown for a similar material --CGT, which possesses a much lower perpendicular magnetic anisotropy (PMA)~\cite{sun2021ultra}-- in a similar experimental configuration, we do not observe any signs of magnetization precession in our measurements.
We explain the absence of the FMR mode in our experiments (figure\,\ref{fig:4}a) by the weak interlayer exchange coupling $A_{\text{ex},\perp}$ due to the vdW stacking in FGT~\cite{van2020layer, meijer2020chiral, wang2020pressure}.
Furthermore, our EDX data (See section\, IV of the supplementary information~\cite{supplement}) indicate that in-plane anisotropy fluctuations within the probes area ($\sim 3$ \SI{}{\micro\meter}) are not unexpected.
We hypothesise that these effects lead to a strong dephasing of the measured precessional motion, which could explain the absence of FMR in our measurements.

To test this hypothesis, we use modified LLG equations which take into account the longitudinal magnetization dynamics~\cite{lee2014gilbert}.
Within this framework, the magnetization is given by $\vec{M}=[1-\Delta M(t)]M_\text{S}\vec{m}$, where $\vec{m}$ is the unit magnetization vector and $\Delta M(t)$ is the experimental data measured at zero field.
Because perturbations are large, an extended LLG equation is used, given by:
\begin{align}
\label{LLG}
\dot{\vec{m}}=& \frac{\gamma}{1+\alpha^{2}}(\vec{m} \times \vec{H}_\text{eff}+\alpha[1-\Delta M(t)] \vec{m} \times \vec{m} \times \vec{H}_\text{eff}) \nonumber\\
&+\frac{1}{\left[1-\Delta M(t)\right]} \frac{\text{d} \Delta M(t)}{\text{d} t} \vec{m}\,,
\end{align}
where the last term takes into account the magnitude change of $\vec{M}$. The effective field $H_\text{eff}$ is given by:
\begin{align}
\label{Heff}
\vec{H}_\text{eff}=& \vec{H}_\text{ext} + \vec{H}_\text{demag}+ \vec{H}_\text{ani} + \vec{H}_\text{ex}\nonumber \\
=&\vec{H}_\text{ext} - \tilde{N} \cdot \vec{M} +\frac{2 K_\text{U}}{\mu_0 M_\text{S}} M_\text{z} \nonumber \\
&+ \frac{2 A_{\text{ex},\perp}}{\mu_0 M_\text{S}} \sum_{i}\frac{\vec{M}-\vec{M_i}}{d_i^2},
\end{align}

\noindent where $\tilde{N}$ is the demagnetizing tensor, with $N_\text{zz}=1$ for thin films.
We simulate every individual layer as a single macrospin coupled to its neighbouring layers $i$ (with magnetization $\vec{M_i}$ and at a distance $d_\text{i}$) via an exchange field, of which the strength is given by the interlayer exchange stiffness $A_{\text{ex},\perp}$.
To be able to compare experiments and simulations reliably, laser attenuation is taken into account.
This is done by scaling the excitation energy via $\Delta M(t)$ with an exponential decay function characterized by the laser penetration depth $\lambda$. This scaling also applies as weights when averaging the magnetization of the layers to simulate laser probing.

Our simulations indicate a significant decrease of the amplitude of the FMR mode for weaker interlayer coupling, caused by a strong dephasing of the magnetization oft the various layers throughout the flake.
Assuming a maximal Gaussian in-plane anisotropy variation $\sigma_\text{K}$ of 5\% and using the parameters presented in section\,VIII of the supplementary information, we simulate the FMR amplitude as a function of the reduced interlayer exchange stiffness.
In figure\,\ref{fig:4}e, we compare this amplitude to the noise threshold, which we estimate from experimental data at negative time delay.
The simulations indicate a significant decrease of the FMR mode amplitude for weaker interlayer coupling, crossing the noise threshold indicated by the light-blue area at around $A_{\text{ex},\perp}$/$A_{\text{ex},\newparallel}=0.2$, where $A_{\text{ex},\newparallel}$ is the in-plane exchange interaction.
This gives us an upper limit for the interlayer exchange stiffness, which is in line with first principle calculations~\cite{wang2020pressure} as well as other experimental observations~\cite{meijer2020chiral}, where the ratio between the interlayer and intralayer exchange parameter is reported to be in the order of 0.1.

Although FMR oscillations were too small to measure, we note that any FMR signal superimposed onto the demagnetization can in principle lead to a shift of the minimum of the demagnetisation curve.
We model this field dependence using equation\,\ref{LLG} and plot the results in figure\,\ref{fig:4}b with a green line.
Although a slight field dependence is calculated, it is an order of magnitude smaller than what we observe in our experiments.
Therefore, we believe that the anisotropy of the spin-flip probability is the only viable explanation for the observed decrease of $t_\text{min(m)}$ as a function of applied field.

\section{Conclusion and Outlook}
\noindent Our results on magnetization dynamics in FGT shine light on the use of ultrafast demagnetization to obtain important magnetic parameters in vdW systems.
Even though a two-step demagnetization process was obtained, which is usually assigned to a long spin lifetime, a detailed analysis of our results point to the opposite, showing that care must be taken when analysing such results.
The strong anisotropy of the spin-flip rate with respect to the magnetization direction, as obtained by our M3TM analysis, provides a powerful method to tune the spin dynamics in two-dimensional magnets.

This property could be further explored to tune the temporal profile of laser-induced spin-currents in vdW magnets~\cite{Choi2014SpinDemagnetization}, or to control the spin and magnon transmission in vdW magnets through a local change of the magnetic anisotropy by electrical gating. 
Furthermore, the slow dynamics observed in FGT together with its unique magnetic properties make it a prime candidate for all-optical magnetization switching based on vdW materials, when interfaces with a material which exhibits fast dynamics~\cite{lalieu2017deterministic}. 
This would open up new pathways towards ultrafast control of magnetization in future data storage devices.

\medskip
\section{Acknowledgements}
We thank Mark C.H. de Jong his help with AFM measurements, and Jeroen Francke, Bart van Looij and Gerrie Baselmans for technical support. This work is supported by Stichting voor Fundamenteel Onderzoek der Materie (FOM) through grant 10023746 and the Nederlandse Organisatie voor Wetenschappelijk Onderzoek (NWO) through grant 10018479. MHDG acknowledges NWO for financial support through the grant Veni 15093.

\section{Supplementary Material}
\beginsupplement

\subsection{More details on pump-probe spectroscopy and the experimental setup}

\noindent The magnetization of the FGT flake is probed using the magneto-optical Kerr effect (MOKE), which is a second-order non-linear optical effect that couples the electric field (e.g. polarization direction) of an incident laser beam to the magnetization of the material it reflects off of. A change in magnetization can thus be measured by carefully probing the polarization of the reflected light, which is done with a crossed polarized scheme. \\

\noindent Ultrafast laser-pulse induced dynamics are measured using pump-probe spectroscopy, where the time-resolved magnetization is probed using MOKE. Linearly polarized laser pulses are generated by a Ti:sapphire laser system at a repetition rate of 82 MHz and a wavelength of 735 nm. The pulse duration at sample position is in the order of 85 fs. Both pump and probe beams are focused onto the flake with a spots of size of $9.6\times 6.6$ \SI{}{\micro\metre\squared} and $2.9\times 1.7$ \SI{}{\micro\metre\squared} respectively. To extract the pump-induced change of the magnetization, the pump beam is mechanically chopped at 67 Hz. Furthermore the polarization of the probe is modulated at 50 kHz by a photoelastic modulator, where the optical axis is chosen such that the polarization oscillates between linearly and circularly polarized light. A dual lock-in scheme, using the two aforementioned frequencies, is used to measure the magneto-optical (MO) signal. The setup geometry is chosen such that only the out-of-plane component of the Kerr rotation is probed. A detailed description of the technique and a schematic overview of the used setup can be found in refs.\,\cite{koopmans2000femtosecond} and \cite{guimaraes2018spin}.\\
\newpage

\subsection{Data normalization}

\begin{figure}[h!]
\includegraphics[width=0.45\textwidth]{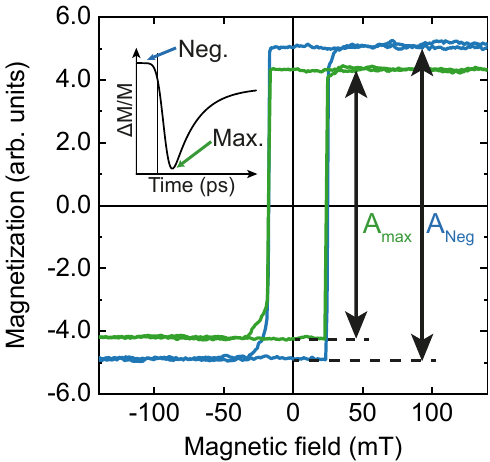}
\caption{\label{fig:5A} Hysteresis loops typically measured to normalize demagnetization data. As indicated in the inset, loops are measured at negative delay and maximum demagnetization. The ratio between the two stepsizes indicates the lost angular momentum in the probed sample area.}
\end{figure}

\noindent The demagnetization curves presented in this paper are normalized by comparing the stepsize of hysteresis loops at negative time delay (when the probe arrives before the pump) and the delay at maximum demagnetization (figure\,\ref{fig:5A}). The extracted value $1-A_\text{Neg}$/$A_\text{Max}$ indicates the percentage of the lost magnetization during demagnetization in the probed area. In a similar fashion, by comparing hysteresis loops measured at negative time delay and without the pump beam, the remaining magnetization value at temporal pump-probe overlap is deduced. An average of three loops is taken to determine stepsizes accurately. 

\noindent The measured value of interest is the maximum demagnetization at peak laser power, measured at the centre of the spatially Gaussian laser pulse. However, due to the similar pump ($\sigma_{pu}$) and probe ($\sigma_{pr}$) size, an average over a much larger region is measured.  The ratio between the measurements and the maximum demagnetization is given by:
\begin{equation}
\frac{\Delta m_{0}}{\Delta m_{\max }}=\frac{\sigma_{pu, x} \sigma_{pu, y}}{\sqrt{\left(\sigma_{pr, x}^{2}+\sigma_{pu, x}^{2}\right)\left(\sigma_{pr, y}^{2}+\sigma_{pu, y}^{2}\right)}}\, .
\end{equation}
This equation is used to extract the maximal demagnetization $\Delta m_{\max }$ using the experimentally observed demagnetization averaged over the probe beam area $\Delta m_{0}$.

\newpage

\subsection{Data fitting using the 3-temperature model}
\noindent We employ the phenomenological 3-temperature model to extract the demagnetization timescales from our measurements. Here, the electron, phonon and spin baths are coupled via phenomenological coupling parameters. Assuming an instantaneous rise of the electron temperature upon laser excitation and negligible spin specific heat, a fitting function that describes laser-induced demagnetization can be derived is the low-fluence limit \cite{dalla2007influence}. This function is given by 
\begin{equation}
\begin{aligned}
-\frac{\Delta M_\text{z}(t)}{M_\text{s}}=& {\left[\left(A_{1} F\left(\tau_\text{d}, t\right)-\frac{\left(A_{2} \tau_\text{e}-A_{1} \tau_\text{m}\right) e^{-\frac{t}{\tau_\text{m}}}}{\tau_\text{e}-\tau_\text{m}}\right.\right.} \\
&\left.\left.-\frac{\tau_{e}\left(A_{1}-A_{2}\right) e^{-\frac{t}{\tau_\text{e}}}}{\tau_\text{e}-\tau_\text{m}}\right) \Theta(t)+A_{3} \delta(t)\right]\\
& \star \Gamma(P_0,\sigma,t)\, .
\end{aligned}
\end{equation}
Here, $A_1$ represents the magnetization value after temporal equilibrium between the baths. $A_2$ is a measure for the initial rise of the electron temperature and $A_3$ captures non-magnetic effects that occur during temporal pump-probe overlap. $F\left(\tau_\text{d}, t\right)$ describes sample cooling via heat diffusion. In our case we use a simple decaying exponential function, where $\tau_\text{d} \gg \tau_\text{e}, \tau_\text{m}$. $\Theta(t)$ is the Heaviside step-function used to describe laser heating and $\delta(t)$ a delta function. A convolution with a Gaussian temporal laser profile $\Gamma(P_0,\sigma,t_0)$ is taken, where $P_0$, $\sigma$ and $t_0$ represent the laser fluence, pulse length and time of excitation respectively.\\

\noindent To avoid overparameterization of the fitting function, several parameters are shared for all measurements. The pulse width $s=0.085$ ps is fixed and the heat diffusion constant, which is determined by material parameters, are shared among all measurements. As mentioned in the main text, $\tau_\text{e}$ is assumed to be temperature independent, and can therefore be shared as well. 
\newpage

\subsection{Fe content in bulk FGT crystal}

\begin{figure}[h!]
\includegraphics[width=0.85\textwidth]{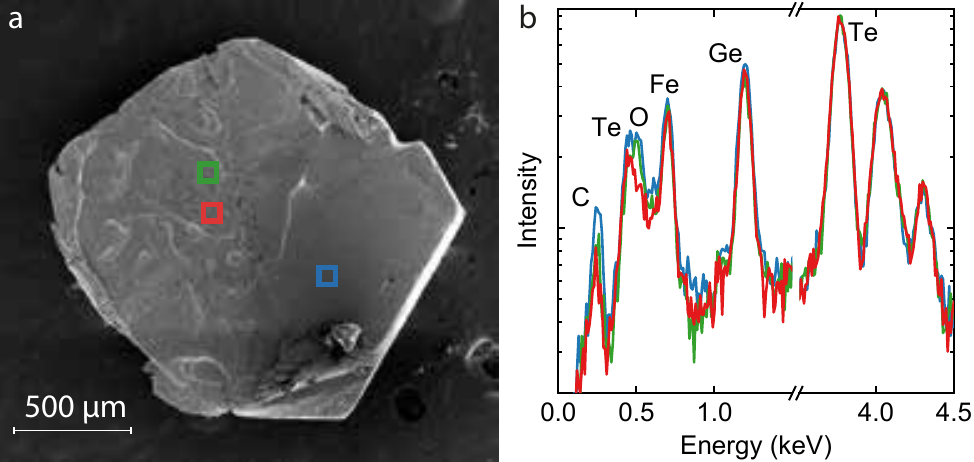}
\caption{\label{fig:1A} (a) SEM image of the bulk FGT crystal used for EDX measurements. The colored boxes indicate areas where the measurements shown in (b) were performed.}
\end{figure}

\noindent We performed energy-dispersive X-ray spectroscopy (EDX) measurements to verify the Fe content in our bulk FGT crystal. The scanning electron microscope image is shown in figure\,\ref{fig:1A}a and the spectra corresponding to the indicated squares in figure\,\ref{fig:1A}b. The resulting composition was normalized on the abundance of tellurium. This approach is validated by considering that the abundance of Te was mostly constant over the different measurements, with an average of $2.00 \pm 0.03$, where the standard deviation is attributed to the measurement error. We measure $\text{Fe}_{2.73(4)}\text{Ge}_{1.17(2)}\text{Te}_{2}$, $\text{Fe}_{2.99(4)}\text{Ge}_{1.04(1)}\text{Te}_{2}$ and $\text{Fe}_{2.94(4)}\text{Ge}_{1.11(2)}\text{Te}_{2}$ for the blue, red and green rectangle respectively. As follows, the crystal is iron deficient, explaining the deviation of $T_C$ from its literature value. In addition, the deficiency is found the be non-uniform across the crystal, which can lead to spatial variations in its magnetic properties.

\newpage
\subsection{Atomic force microscopy image of the FGT flake shown in the main text}
\begin{figure}[h!]
\includegraphics[width=0.85\textwidth]{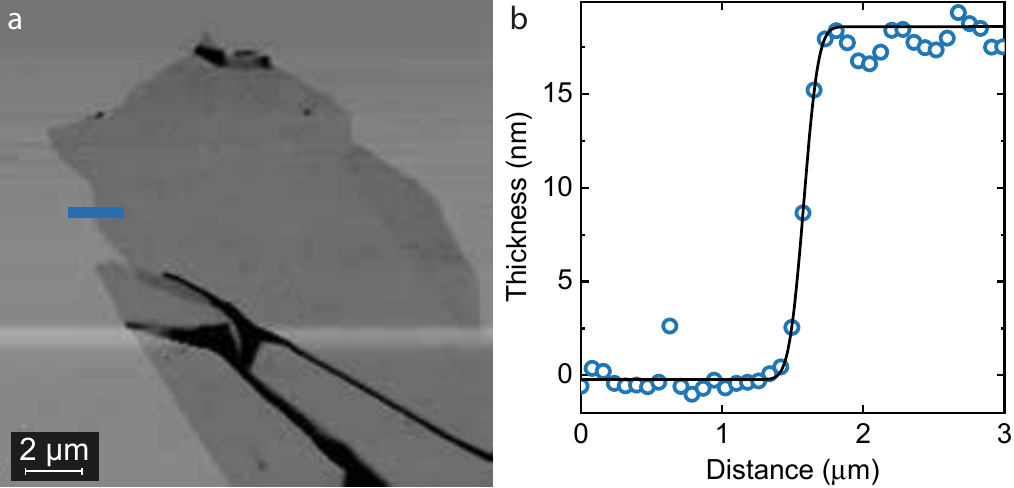}
\caption{\label{fig:2A} (a) AFM image of our FGT flake used in the main text, zoomed in on the area probed in our experiments. The blue line indicates the thickness scan shown in (b).}
\end{figure}

\noindent Atomic force microscopy (AFM) is used to determine the thickness of our FGT flake (figure\,\ref{fig:2A}). The results indeed show the probed area is uniform in thickness. We extract the thickness at 5 different edges, yielding an average thickness of $16.8 \pm 0.7$ nm, corresponding to approximately 20 layers when assuming $\sim\!0.8$ nm layer spacing.

\newpage

\subsection{Full datasets of ultrafast demagnetization}
\begin{figure}[h!]
\includegraphics[width=0.9\textwidth]{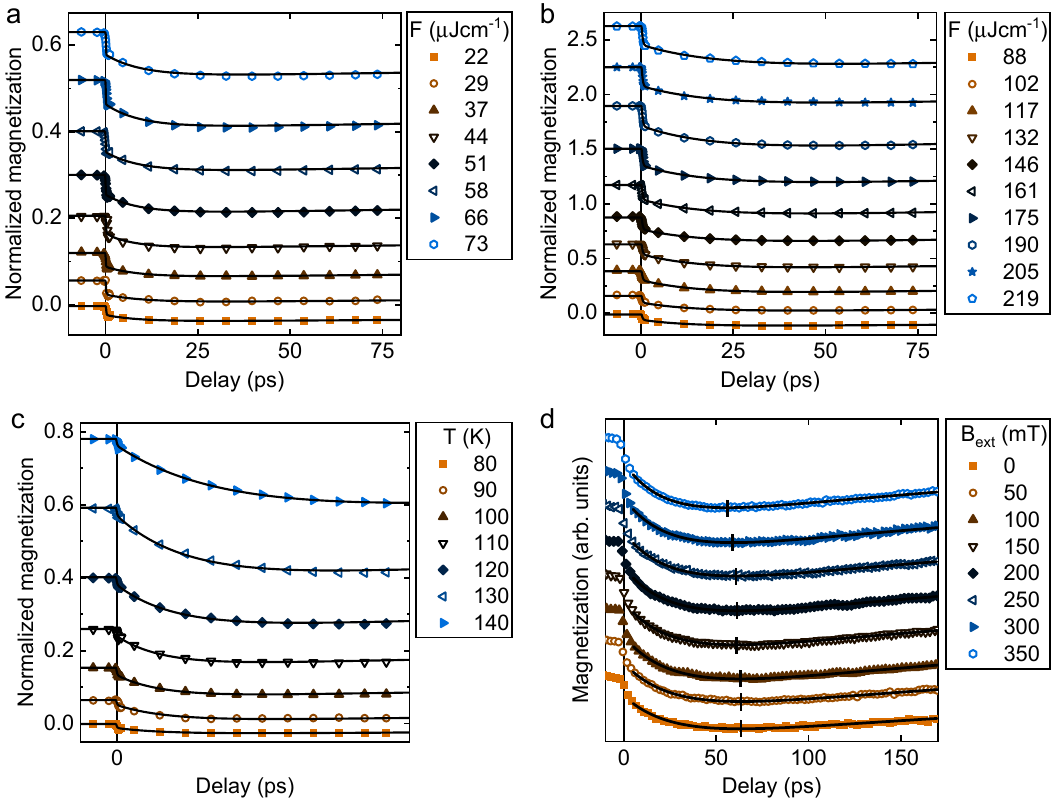}
\caption{\label{fig:4A} Additional experimental data ultrafast demagnetization. (a) and (b) show the full data set for the fluence dependence shown in figure\,2 of the main text. (c) Full temperature dependence shown in figure\,2d of the main text. (d) Full field dependence shown in figure\,4b of the main text. Black lines indicate fits as described in the main text.}
\end{figure}

\noindent For clarity, only a few data sets are shown in the main text. In figure\,\ref{fig:4A}, all experimental data are presented. 

\newpage
\subsection{More details on the microscopic 3-temperature model}

\noindent In the microscopic 3-temperature model, the energy exchange between electron and phonon baths is governed by the 2-temperature model:
\begin{align}
\label{eq1a}
\gamma T_\text{e} \dfrac{\text{d} T_\text{e}}{\text{d} t} =& g_\text{ep} (T_\text{p}- T_\text{e})+\frac{P_0}{\sigma \sqrt{\pi}}e^{\frac{-t^2}{\sigma^2}}\, ,\\
\label{eq1b}
C_\text{p} \dfrac{\text{d} T_\text{p}}{\text{d} t} =& g_\text{ep} (T_\text{e}- T_\text{p}) + C_\text{p}\frac{T_\text{amb}-T_\text{p}}{\tau_\text{d}}\, .
\end{align} 
Here, $\gamma$, $C_\text{p}$ and $g_\text{ep}$ the electron heat capacity constant, the phonon heat capacity and the electron-phonon interaction constant respectively. The electron system is heated by the laser pulse, which is characterized by absorbed laser fluence $P_0$ and pulse duration $\sigma$. Heat diffusion with a typical timescale $\tau_\text{d}$ from the phonon system to the surrounding ($T_\text{amb}$) is included as well. The above equations are used to simulate the data in figure\,3a in the main text.

Equation\,3 of the main text was first derived by Schellekens \cite{schellekens2014manipulating}. While ignoring heat dissipation in the the FGT flake, for small perturbations and near $T = T_\text{C}$, the temperature dependence of the magnetization can be written as: 
\begin{align}
\label{B1}
    m=\sqrt{3\left(1-\frac{T}{T_{C}}\right)}.
\end{align}
The temperature dependence of the magnetic relaxation timescale can be determined by assuming that the electron and phonon baths are heated instantly by an amount $\Delta T$. In this case, $\tau_\text{m}$ is approximated by
\begin{align}
\label{B2}
    \tau_\text{m}=\frac{\Delta m(\infty)}{\dot m (0)}\,.
\end{align}
The total lost magnetization can be calculated using equation\,\ref{B1}:
\begin{align}
\label{B3}
    \Delta m(\infty)=\Delta T \left.\frac{\text{d}m}{\text{d}t}\right\vert_{T=T_\text{t=0}} =\frac{\sqrt{3}\Delta T}{2\sqrt{T_\text{C}(T_\text{t=0}-T_\text{C})}}
    \,.
\end{align}
Expanding equation\,1 of the main text around $m=0$ and assuming an equilibrated system at $T_\text{e}=T_\text{p}=T$ yields
\begin{align}
\label{B4}
    \dot m(0)=\Delta T \left.\frac{\text{d}\dot m}{\text{d}t}\right\vert_{T=T_\text{t=0}} = -\frac{\Delta T R m}{T_\text{C}}\,.
\end{align}
Now we can substitute equation\,\ref{B3} and \ref{B4} into equation\,\ref{B2} to calculate $\tau_\text{m}$, yielding
\begin{align}
    \tau_\text{m}=\left(\frac{1}{2R}\right)\left(\frac{1}{1-\frac{T_\text{t=0}}{T_\text{C}}}\right)\,.
\end{align}
If the magnetic material is allowed to completely relax back to equilibrium between laser excitation events, $T_\text{t=0}$ would be equal to the ambient temperature. However, between two consecutive pulses the FGT has not been completely remagnetized, which leads the an average heating effect of the flake, elevating it's average temperature above the ambient temperature. This effect is solely induced by the laser fluence and is therefore temperature independent. This means $T_\text{t=0}$ can be written as $T +\Delta T_\text{h}$, where $T_\text{h}$ is the temperature rise due to the average heating effect.

\newpage

\subsection{Simulation parameters}

\begin{table*}[h!]
\centering
\caption{\label{tab:tab1} Numerical values used in the microscopic 3-temperature model (M3TM) and Landau–Lifshitz–Gilbert (LLG) calculations. The values for the respective models are separated by a horizontal line.}
\begin{tabular}[t]{llccc}
\hline
Model & symbol &  meaning & estimate & unit \\
\hline
M3TM & $\mu_at$ & Atomic magnetic moment \footnotemark[1]& $ 1.47\times 10^{-23} $ & $\mbox{Am}^{2} $ \\ 
 & $\gamma$ & Electron heat capacity constant \footnotemark[1]& $ 1561 $ & $\text{Jm}^{-3}\text{K}^{-2}$ \\ 
 & $V_\text{at}$ & Atom spin density volume \footnotemark[1]& $1.68\times 10^{-29}$ & $\mbox{m}^{3} $\\
 & $T_\text{C}$ & Curie temperature \footnotemark[2]& 191 & \text{K} \\
&$\sigma$ & pulse width \footnotemark[2]& 85 & \text{fs} \\
&$g_\text{ep}$ & Electron-phonon relaxation rate \footnotemark[3]& $ 1.33\times 10^6 $ &  $\text{Jm}^{-3}\text{K}^{-1}\text{ps}^{-1}$ \\
&$C_\text{p}$ & phonon heat capacity \footnotemark[3]&  $6.28 \times 10^6$ & $\text{Jm}^{-3}\text{K}^{-1}$ \\
&$E_\text{D}$ & Debeye energy \footnotemark[3]&  $6.49 \times 10^{-21}$ & J \\
&$a_\text{sf}$ & Spin-flip probability & 0.14 & \text{unity}\\
&$\tau_\text{d}$ & Heat diffusion timescale & 350 & ps \\
\hline
LLG&$K_\text{U}$ & Uniaxial anisotropy constant \footnotemark[4] & 438 & kJm$^{-3}$ \\
&$M_\text{S}$ & Saturation magnetization \footnotemark[4] & 225.6 & kAm$^{-1}$ \\
&$A_\text{ex,//}$ & In-plane exchange stiffness \footnotemark[5] & 1 &  pJm$^{-1}$ \\
&$d$ & Interlayer distance \footnotemark[6] & 0.817 & nm \\
&$\lambda$ & Laser penetration depth \footnotemark[7] & 20 & nm \\
&$\alpha$ & damping parameter \footnotemark[7] & 0.05 & unity \\
\hline
\footnotetext{Taken from Ref.\,\cite{zhu2016electronic}.} 
\footnotetext{Based on measurements presented in main text.}
\footnotetext{First estimations based on Fe \cite{sultan2012electron}.}
\footnotetext{Estimated for $T=160$ K based on measurements done in Ref.\,\cite{leon2016magnetic} and \cite{guo2021temperature}.}
\footnotetext{Taken from Ref.\,\cite{leon2016magnetic}.}
\footnotetext{Taken from Ref.\,\cite{villars2014handbook}.}
\footnotetext{Estimated.}
\end{tabular}
\end{table*}%

\subsection{Relation between the maximum demagnetization and the spin-flip probability}
\begin{figure}[h!]
\includegraphics[width=0.5\textwidth]{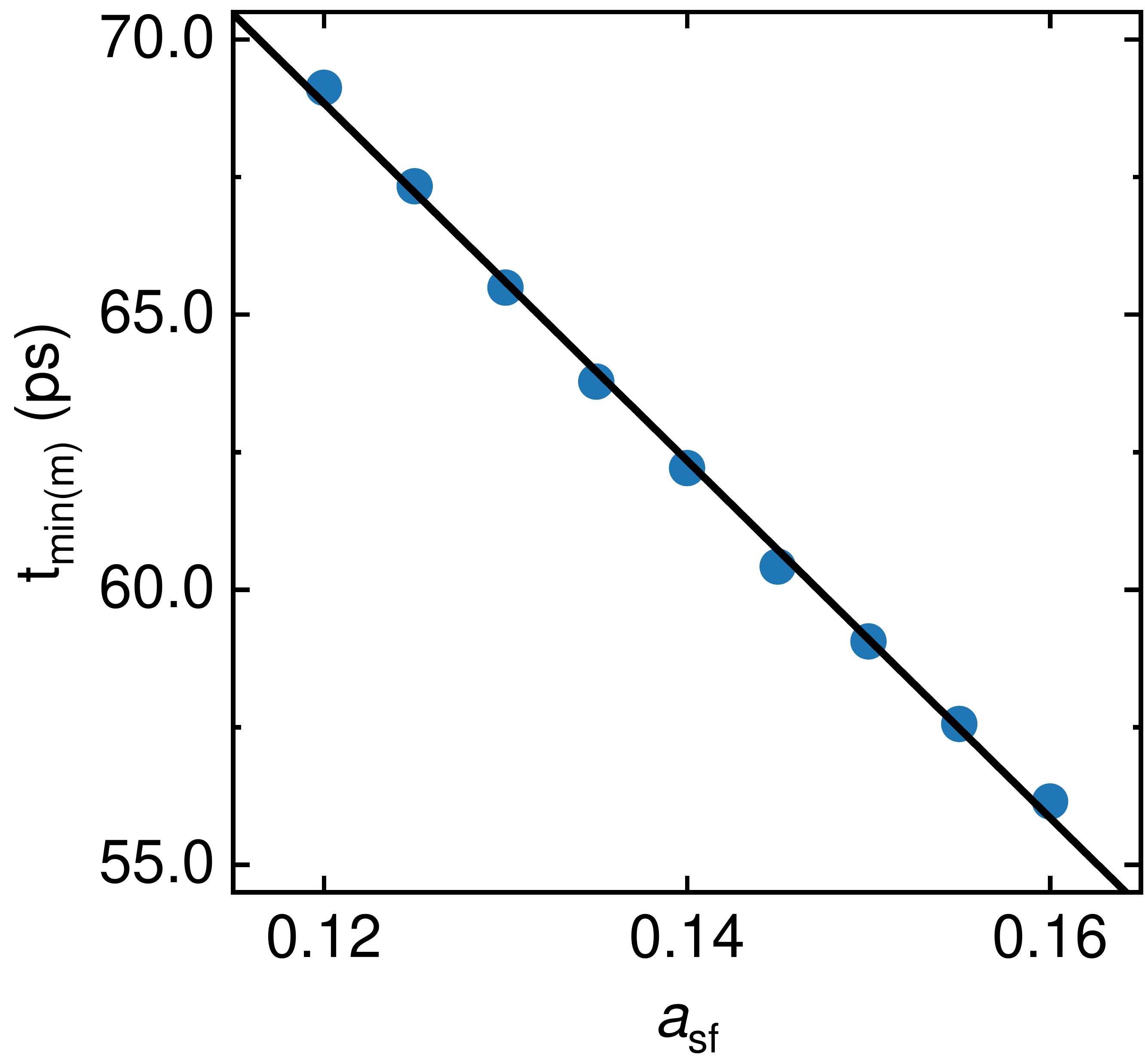}
\caption{\label{fig:3A} The relation between the time delay for maximum demagnetization and the spin-flip probability, as calculated with the M3TM.}
\end{figure}

\noindent We employed the M3TM to find the relation between the delay at the minimum of the demagnetization trace and the spin-flip probability $a_\text{sf}$ (figure\,\ref{fig:3A}). This delay $t_\text{min(m)}$ linearly decreases as $a_\text{sf}$ is increased. A linear fit of the data yields:
\begin{align}
    t_\text{min(m)} = 107.8 -324.7 \cdot a_\text{sf}\,.
\end{align}

\newpage

\subsection{Average laser heating of the FGT flake}
\begin{figure}[h!]
\includegraphics[width=0.5\textwidth]{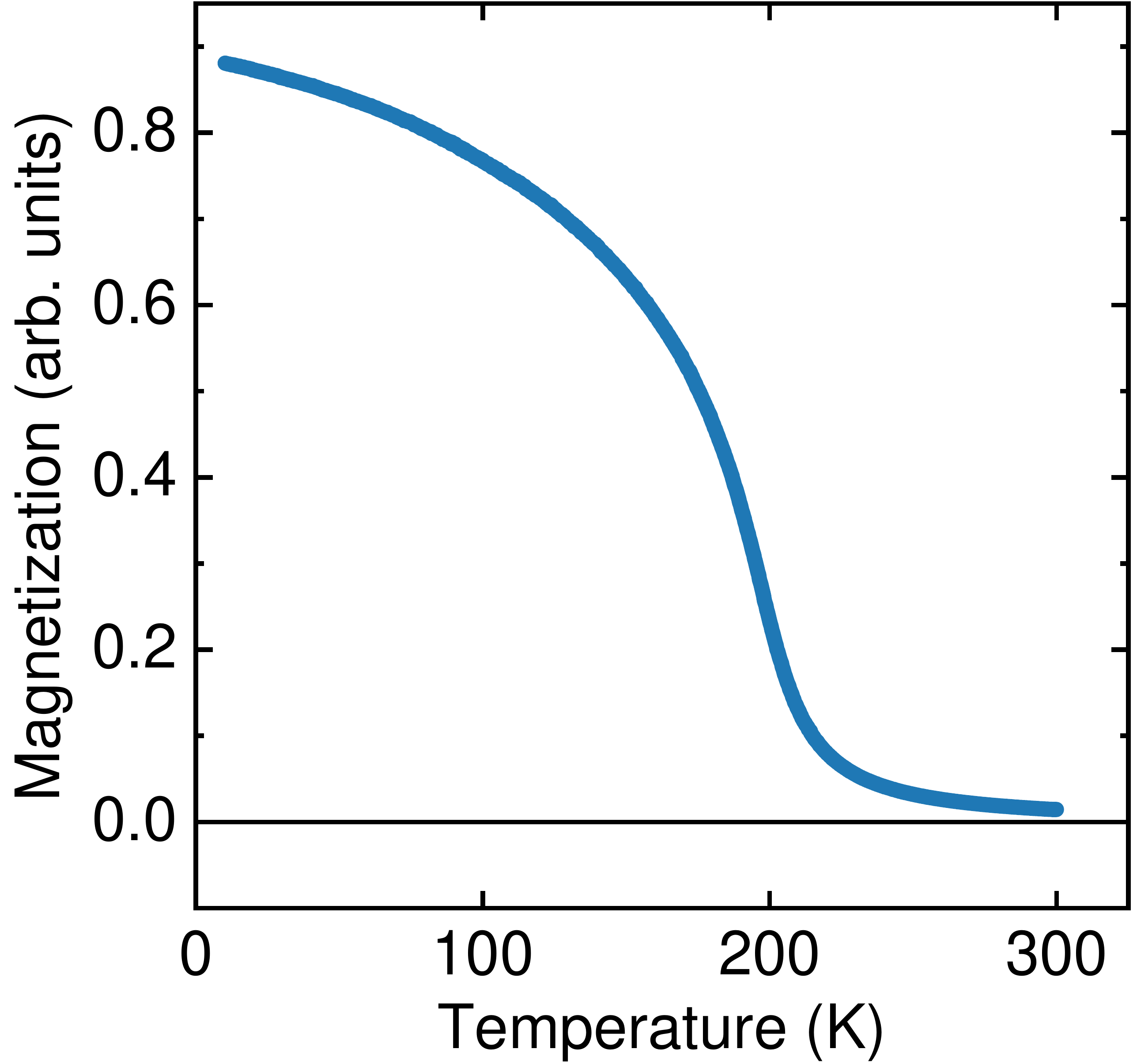}
\caption{\label{last} Bulk magnetization as a function of temperature measured using vibrating sample magnetometry.}
\end{figure}

\noindent This temperature difference $\Delta T_\text{h}$ due to laser heating can also be estimated based on the step size of hysteresis loops following a similar procedure as described in Sec.\,II of this Supplementary Information. 
Here, hysteresis loops taken at negative time delay are compared, measured with and without sample pumping. 
In this way, the impact of average heating on the magnetization is measured.
The relative difference in magnetization due to average heating can then be related to temperature rise using the experimentally obtained $M$ vs. $T$ (vibrating sample magnetometry) curve for bulk FGT as as a look-up table (figure\,\ref{last}).
From this we estimate an average heating of $\Delta T_\text{h} = 16 \pm 7$, in line with the fitting parameter found in the main text. 

\clearpage
\bibliography{Main}
\end{document}